\documentclass[a4paper,twocolumn,11pt,accepted=2025-03-31]{quantumarticle}
\pdfoutput=1
\usepackage{graphicx}
\usepackage{dcolumn}
\usepackage{bm}
\usepackage{booktabs}
\usepackage{braket}
\usepackage{amsmath}
\usepackage{subcaption}
\usepackage{xcolor}
\usepackage{multirow}
\usepackage{soul}
\usepackage{hyperref}
\hypersetup{breaklinks=true}
\usepackage[numbers]{natbib}
\begin{document}
\newcommand{\ct}{\hat{a}^\dag}
\newcommand{\an}{\hat{a}}
\newcommand{\bvac}{\bra{\text{vac}}}
\newcommand{\kvac}{\ket{\text{vac}}}

\title{Efficient High-Dimensional Entangled State Analyzer with Linear Optics}

\author{Niv Bharos}
\affiliation{QuTech, Delft University of Technology, Delft 2628CJ, Zuid-Holland, The Netherlands
}
\author{Liubov Markovich}
\affiliation{
Instituut-Lorentz, Universiteit Leiden, P.O. Box 9506, 2300 RA Leiden, The Netherlands}
\affiliation{$\langle \text{aQa}^\text{L} \rangle$ Applied Quantum Algorithms Leiden, The Netherlands}
\author{Johannes Borregaard}
\affiliation{QuTech, Delft University of Technology, Delft 2628CJ, Zuid-Holland, The Netherlands
}
\affiliation{Department of Physics, Harvard University, Cambridge, Massachusetts 02138, USA}

\begin{abstract}
The use of higher-dimensional photonic encodings (qudits) instead of two-dimensional encodings (qubits) can improve the loss tolerance and reduce the computational resources of photonic-based quantum information processing. To harness this potential, efficient schemes for entangling operations such as the high-dimensional generalization of a linear optics Bell measurement will be required. We show how an efficient high-dimensional entangled state analyzer can be implemented with a linear optics interferometer and auxiliary photonic states. The degree of entanglement of the auxiliary state is much less than in previous protocols as quantified by an exponentially smaller Schmidt rank. In addition, the auxiliary state only occupies a single spatial mode, allowing it to be generated deterministically from a single quantum emitter coupled to a small qubit register. The reduced complexity of the auxiliary states results in a high robustness to imperfections and we show that auxiliary states with fidelities above $0.9$ for qudit dimensions $4$ can be generated in the presence of qubit error rates on the order of $10\%$. This paves the way for experimental demonstrations with current hardware.

\end{abstract}

\maketitle

\section{Introduction}
\par 
Encoding quantum information in photonic degrees of freedom is at the heart of both quantum networking~\cite{Kimble2008} and photonic-based quantum computing~\cite{Slussarenko2019}. In the former, photonic qubits can be used to entangle distant stationary qubit systems~\cite{Moehring2007,Hensen2015,Leent2022}, while in the latter, multi-photon entangled states can serve as a resource for universal quantum computing with linear optics~\cite{bartolucci2023fusion, Browne2005,Istrati2020}. Both approaches take advantage of the excellent coherence properties of single photons and the easy manipulation of photonic qubits through linear optical elements such as phase shifters and beam splitters. 
\par
Photons possess multiple degrees of freedom, such as polarization, frequency, the time-bin basis or spatial modes. The dimensionality of photons can be increased by encoding the quantum state in multiple $(d)$ orthogonal modes, which creates a photonic qudit that provides a higher information density than photonic qubits: a $d$-dimensional photonic state encodes up to $\lfloor \text{log}_2{(d)}\rfloor$ qubits of information~\cite{Knall2022,Borghi2023}.
\par
Photonic qudits have several advantages compared to photonic qubits. For entanglement generation recent work has shown how photonic qudits lower the quantum memory requirements for the generation of multiple high-fidelity entangled pairs~\cite{Piparo2019,zheng2022entanglement,xie2021quantum, zhou2023parallel}. In quantum cryptography, photonic qudits enable more error-robust quantum key distribution~\cite{yang2021feasible_MDIQKD,erhard2020advances,cerf2002security,bechmann2000quantum,Bell2022}. For photonic-based quantum computing, qudit encodings can provide means for more efficient algorithms with reduced circuit depth~\cite{zhou2003quantum,Imany2019,paesani2021scheme}, lower the resource requirements for simulation of high-dimensional gauge theories~\cite{meth2023, gonzalez2022hardware, meth2023}, and optimization problems~\cite{karácsony2023}. More fundamentally, qudit entanglement can exhibit stronger non-classical correlations than qubit entanglement~\cite{kaszlikowski2000violations, collins2002bell}. 
\par 
Experiments with high-dimensionally entangled photons have attracted much attention in recent years~\cite{kues2017chip, dada2011experimental, schaeff2015experimental, ren2017ground, krenn2014generation, molina2004triggered, hu2020efficient, giovannini2013characterization,agnew2011tomography, erhard2018experimental,wang2020qudits}. The crux of all photonic-based quantum information processing is the ability to perform entangling operations between photonic systems. For photonic qubits, this can be obtained with linear optics through probabilistic Bell measurements~\cite{knill2001scheme,Kok2007,calsamiglia2001maximum}. In order to harness the full potential of photonic qudits, similar high-dimensional techniques have to be devised. In three dimensions, the success rates of existing protocols are high enough to perform experiments, as has been shown with the quantum teleportation of a three-dimensional state~\cite{hu2020experimental,luo2019quantum}. However, devising a scheme for a linear-optics, high-dimensional entangled state analyzer (ESA) with success rates that scale efficiently as the photonic dimension increases has proven to be difficult and remains an outstanding problem in the field.
\par 
Auxiliary photonic states are necessary to perform an ESA~\cite{calsamiglia2002generalized}. Specifically, an ESA in $d$ dimensions requires at least $d-2$ extra photons. Protocols with non-entangled auxiliary states experience an exponentially decreasing success probability as the dimension grows~\cite{luo2019quantum}, which severely limits their feasibility. The use of entangled auxiliary states allows, in principle, for efficient operation where the success probability only decreases quadratically with the dimension~\cite{zhang2019quantum}. The protocol in Ref.~\cite{zhang2019quantum}, however, requires the Schmidt rank of the auxiliary state to increase faster than exponentially with the dimension of the input states. Generating such highly complex states is challenging, and known generation protocols only succeed with exponentially decreasing success probability~\cite{zhang2019quantum}.
\par 
In this paper, we introduce a novel protocol for linear optics ESA in even dimensions, that circumvents the exponential scalings of previous proposals. Our protocol achieves an efficient $d$-dimensional ESA with success probability $2/d^2$ for  high-dimensional entanglement swapping using an auxiliary entangled state with Schmidt rank scaling as $d/2$. Furthermore, we outline how the auxiliary states can be generated deterministically from a single quantum emitter coupled to a small qubit processor with $\sim\log{(d/2)}$ qubits. We show that high-fidelity auxiliary states with fidelity $>0.9$ can be generated with our protocol even in the presence of noise rates on the order of 10\% per gate operation. Consequently, our protocol is compatible with current quantum hardware such as solid-state defect centers~\cite{Knall2022,Meng2023,pompili2021} or atomic qubits~\cite{Thomas2022} and outlines a feasible route towards efficient high-dimensional photonic quantum information processing. 
\section{High-Dimensional Entangled State Analyzer}
We consider time-bin encoded qudits, where the presence of a single photon in different time slots, or `bins', is used to encode quantum information. Although we focus on the time-bin encoding, our scheme is compatible with other photonic qudit encodings such as spatial or frequency encoding. The implementations of the necessary operations of our scheme that we describe here, such as the quantum Fourier transform (QFT) and the generation of the auxiliary state are, however, specific for time-bin encoding. We leave the design and implementation of other encodings to future work.
\par
\begin{figure}[tb]
    \includegraphics[width=\linewidth]{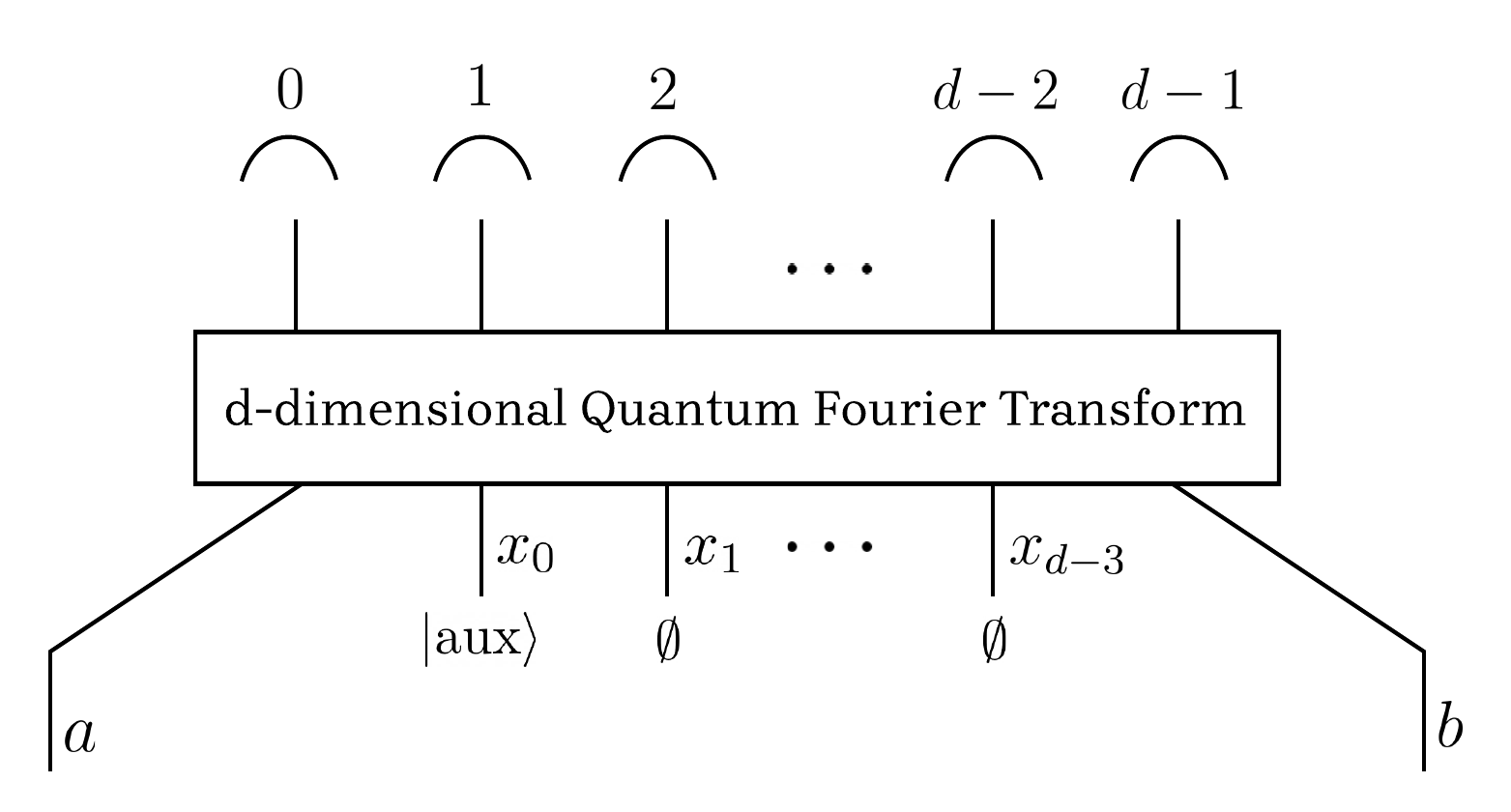}
    \caption{Overview of a $d$-dimensional entangled state analyzer (ESA) of time-bin encoded qudits in spatial modes $a$ and $b$. The protocol uses $d-2$ auxiliary photons in one input mode $x_0$ to the quantum Fourier transform (QFT), the other input modes $x_1, x_2, \ldots, x_{d-3}$ contain vacuum as indicated by $\emptyset$. The QFT of the spatial modes can be implemented with standard linear optical elements. Finally, the output modes are measured with $d$ single photon detectors labeled $0, 1, \hdots d-1$. The condition for a successful ESA is to measure all photons in different time-bins and the resulting entangled state projection is determined by the specific detection pattern of the detectors.}
    \label{fig:figure1}
\end{figure}

 A schematic overview of an ESA of two photonic input qudits following our protocol is shown in Fig.~\ref{fig:figure1}. As it is shown in Ref.~\cite{calsamiglia2002generalized}, performing an ESA in $d$ dimensions requires $d$ initial particles. Thus, besides the two input photons ($a$ and $b$), a minimum of $d-2$ auxiliary photons is required. We mix all input photons with a QFT of the spatial modes, which can be achieved with a sequence of beam splitters and phase shifters~\cite{reck1994experimental, barak2007quantum}.
Finally, we herald on detecting all photons in different time-bins: in these cases the QFT erases the which-path information of all photons: all photons are projected into the permutation basis of all time-bins. The photons can be detected by the same detector or by different detectors, which correspond to projections on different maximally entangled states. We note that these states are not necessarily orthogonal to each other and that the full set of projections do not correspond to a complete basis.  
\par
To maximize the success probability of our setup, we design the auxiliary state of $d-2$ photons such that these photons are never in the same time bin since this would lead to an unsuccessful ESA. Moreover, we use the fact that we project all photons in a superposition of modes where all time bins are unique (the permutation basis) to exclude two unique time bins in each mode of the auxiliary state. By doing so, each mode leads to two non-zero terms in the final projection of $a$ and $b$, as we see later on. The auxiliary photons are therefore prepared in an entangled state of the form
\begin{equation}\label{eq:2}
        \ket{\textrm{aux}} = \frac{1}{\sqrt{d/2}}\sum_{j = 0}^{\frac{d}{2} - 1}\ket{\textrm{aux}_{j}},
\end{equation}
where $|\text{aux}_j\rangle=\ct_{\boldsymbol{\tau_j}[0]}\ct_{\boldsymbol{\tau_j}[1]}\hdots\; \ct_{\boldsymbol{\tau_j}[d-3]}\kvac$ is defined in the second-quantization framework. The state above is a sum of terms labeled by $j$ where the time-bins of the photons are denoted by the creation operators $\ct_{\boldsymbol{\tau_j}[k]}$. Here $\boldsymbol{\tau_j}$ is a vector with elements $\boldsymbol{\tau_j}[k]\in[0,d-1]$.
\par
To ensure that each term in $\ket{\textrm{aux}}$ contains $d-2$ photons in different time-bins, we enforce the constraint $(\boldsymbol{\tau_j}[0], \boldsymbol{\tau_j}[1], \dots, \boldsymbol{\tau_j}[d-3],y_j,z_j) \in P[0, 1, ..., d-1]$, i.e., that they correspond to a permutation $P$ of all the $d$ time-bins. Since there are only $d-2$ auxiliary photons and $d$ time-bins, each term $\ket{\text{aux}_j}$ does not contain photons in exactly two time-bins $y_j$ and $z_j$. We enforce that the two time-bins that are excluded in one state $\ket{\textrm{aux}_i}$ are not equal to the two time-bins excluded from any other term $\ket{\textrm{aux}_j}$, i.e., $y_i\neq y_j\;\land\; z_i\neq z_j\;\land\; y_i \neq z_j, \quad\forall\; i, j$. From these constraints, it follows that the state in Eq.~(\ref{eq:2}) is an entangled state with Schmidt rank $d/2$.
\par
The freedom in choosing the sets $\{y_j,z_j\}$ for the auxiliary state corresponds to choosing what subset of maximally entangled states the ESA will project onto. One possible choice is to assign all $y_j$ values as odd and all $z_j$ values as even. This will result in a specific subset of entangled state projectors set by the detection pattern. In general, there are $d!/(2^{d/2}(d/2)!)$ ways to choose the set $\{y_j,z_j\}$. Notably, a specific choice of the set $\{y_j,z_j\}$ will not necessarily lead to projections onto states that are orthogonal to those corresponding to a different choice of $\{y_j,z_j\}$. We provide the general form of the projectors for specific choices of the auxiliary state and detection patterns in Appendix A, but we will detail the performance of our proposed ESA below for four-dimensional entanglement swapping as an illustrative example.
\section{Entanglement Swapping in 4D}
We discuss four-dimensional entanglement swapping between remote qudit registers to better understand how the protocol works. In this example, two remote qudit systems are entangled by means of an ESA but in a similar way it applies to e.g. fusion of high-dimensional graph states~\cite{paesani2021scheme}. \par
\begin{figure}[t]
    \centering
    \includegraphics[width=0.3\textwidth]{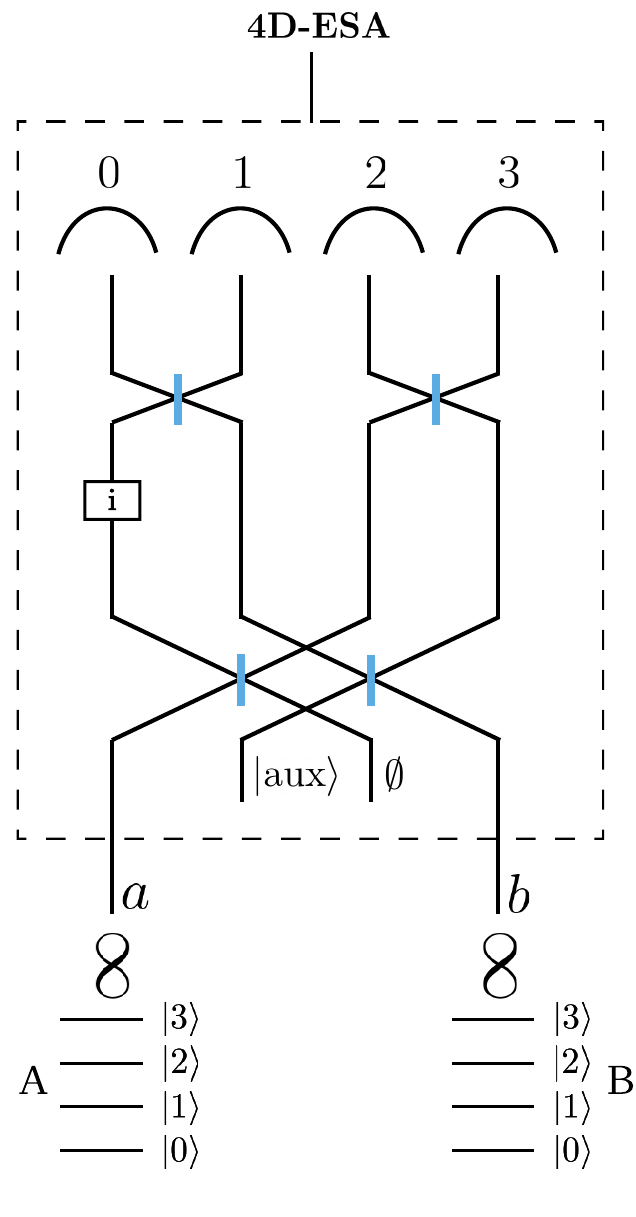}
    \caption{Example of the four-dimensional ESA for the generation of four-dimensional entanglement (equivalent to two Bell pairs) shared by Alice and Bob. The QFT is implemented with four $50$:$50$ beam splitters (blue) and a phase shifter of $\frac{\pi}{2}$.}
    \label{fig:figure2}
\end{figure}

Consider two parties, Alice and Bob, who have entangled their local, stationary qudit systems with photonic qudits. The goal is to entangle their local systems by projecting the photonic states onto a maximally entangled state using a four-dimensional ESA. The details of the setup are shown in Fig.~\ref{fig:figure2}. 
\par The initial state of Alice and Bob's systems is
\begin{equation}\label{eq:3}
     \ket{\psi}_{AaBb} = \frac{1}{4}\sum_{i = 0}^{3}\ct_{i, a}\kvac\ket{i}_A \otimes \sum_{j = 0}^3\ct_{j, b}\kvac\ket{j}_B.
\end{equation}
Here, the modes $A,a$ ($B,b$) belong to Alice's (Bob's) entangled state. We can project the modes $a$ and $b$ onto a 4-dimensional maximally entangled state using an ESA with an auxiliary photonic state of two photons. Without loss of generality, we consider an auxiliary state of the form 
\begin{equation}\label{eq:4}
    \ket{\textrm{aux}} = \frac{1}{\sqrt{2}}\left(\ct_{0, x_0}\ct_{1, x_0} + \ct_{2, x_0}\ct_{3, x_0}\right)\kvac,
\end{equation}
which is a superposition of the two photons being in time bin 0 and 1 or 2 and 3, respectively. Next, we show that this auxiliary state indeed swaps the entanglement. Keeping track of the possible states of all photons as they pass through each linear optical element becomes a large calculation very quickly. Many of the output states will be redundant since we are only interested in certain successful measurement results. Hence we start with conditioning on detecting the four photons (one from both Alice and Bob and two from the auxiliary state) in different time bins and work our way back through the protocol to see the projected state of Alice and Bob's registers. We can express the resulting state of the qudit registers following a successful ESA measurement with a photon detection in each time-bin as
\begin{equation} \label{eq:5}\footnotesize
    \begin{split}
        \ket{\phi}_{AB}=\bvac\an_{0, D_0}\an_{1, D_1}\an_{2, D_2}\an_{3, D_3}\;U_{\text{QFT}}\ket{\psi}_{AaBb}\ket{\textrm{aux}},
    \end{split}
\end{equation}
where $U_{\text{QFT}}$ is the QFT transformation of the spatial modes and the annihilation operator $\an_{i, D_i}$ describes the detection of the output photon in time-bin $i$ at detector $D_i$.
The subscripts $D_i$ with $D_i\in\{0,1,2,3\}$ thus label the spatial mode corresponding to the detector (shown at the top in Fig.~\ref{fig:figure2}) at which a photon in time-bin $i$ is measured. For example, $D_0$ is the detector that clicks in time-bin $0$. As an example, we consider the case where all photons are measured at the first detector: $\small{D_0=D_1=D_2=D_3=0}$. Working backward to find the state $\ket{\phi}_{AB}$ of Alice and Bob, we start with the following projection state:
\\
\begin{equation}\label{eq:6}
    \begin{split}
        \bra{P} = \bvac&\an_{0, 0}\an_{1, 0}\an_{2, 0}\an_{3, 0}\;U_{\text{QFT}}\\
        = \frac{1}{2^4}\bvac\big(&\an_{0, a} + \an_{0, b} + \an_{0, x_0} + \an_{0, x_1}\big)\\
        \otimes\big(&\an_{1, a} + \an_{1, b} + \an_{1, x_0} + \an_{1, x_1}\big)\\
        \quad\;\otimes\big(&\an_{2, a} + \an_{2, b} + \an_{2, x_0} + \an_{2, x_1}\big)\\
        \otimes\big(&\an_{3, a} + \an_{3, b} + \an_{3, x_0} + \an_{3, x_1}\big),        
    \end{split}
\end{equation}
\\
where the spatial mode index of the annihilation operators transforms from the detector modes to the input modes of the ESA. Note that $x_0$ corresponds to the spatial mode at the input of the QFT that contains the auxiliary photons and $x_1$ to the input mode that does not contain any photons. We obtain the expression in the second line by letting $U_{\text{QFT}}$ act to the left. 
\par This projector can be further simplified using the fact that the input state $\ket{\psi}_{AaBb}\ket{\text{aux}}$ has one photon in modes $a$ and $b$ and two photons in mode $x_0$. Thus only terms in the projector that correspond to one photon in both spatial mode $a$ and $b$ and two photons in spatial mode $x_0$ will lead to non-zero overlap. Keeping only the non-zero terms allows us to write the projector as: 
\begin{widetext}
    \begin{equation}
    \begin{split}
        \bra{P'}= \frac{1}{2^4}\bvac\bigg(&\an_{0, a}\an_{1, b}\an_{2, x_0}\an_{3, x_0} + \an_{0, a}\an_{2, b}\an_{1, x_0}\an_{3, x_0} + \an_{0, a}\an_{3, b}\an_{1, x_0}\an_{2, x_0} + \an_{1, a}\an_{0, b}\an_{2, x_0}\an_{3, x_0}\\
        &+ \an_{1, a}\an_{2, b}\an_{0, x_0}\an_{3, x_0} + \an_{1, a}\an_{3, b}\an_{0, x_0}\an_{2, x_0} + \an_{2, a}\an_{0, b}\an_{1, x_0}\an_{3, x_0} + \an_{2, a}\an_{1, b}\an_{0, x_0}\an_{3, x_0}\\
        &+ \an_{2, a}\an_{3, b}\an_{0, x_0}\an_{1, x_0} + \an_{3, a}\an_{0, b}\an_{1, x_0}\an_{2, x_0}+ \an_{3, a}\an_{1, b}\an_{0, x_0}\an_{2, x_0} + \an_{3, a}\an_{2, b}\an_{0, x_0}\an_{1, x_0}\bigg)
    \end{split}
\end{equation}
\end{widetext}
Applying this projection to the input state of the auxiliary photons leaves us with the projection of Alice and Bob's photons $a$ and $b$:
\begin{equation}\label{eq:8}\footnotesize
    \begin{split}
        &\bra{P}_{ab} = \braket{P'\:|\:\textrm{aux}} \\
        &= \frac{1}{2^4\sqrt{2}}\bvac(\an_{0, a}\an_{1, b} + \an_{1, a}\an_{0, b} + \an_{2, a}\an_{3, b} + \an_{3, a}\an_{2, b}).
    \end{split}
\end{equation}
\\
\normalsize
We see that the input modes $a$ and $b$ are projected on a maximally entangled state. Because $a$ and $b$ were entangled with the qudit registers, following a successful ESA, the registers are also projected into an entangled state:\\
\begin{equation} \label{eq:9} \footnotesize
    \begin{split}
        &\ket{\phi}_{AB} = \bra{P}_{ab}\ket{\psi}_{AaBb}\\
        & = \frac{1}{2^5\sqrt{2}}\cdot\frac{1}{2}\left(\ket{0}_A\ket{1}_B + \ket{1}_A\ket{0}_B + \ket{2}_A\ket{3}_B + \ket{3}_A\ket{2}_B\right),
    \end{split}
\end{equation}
\\
corresponding to a (not normalized) maximally entangled state i.e., we have successfully performed a high-dimensional entanglement swap. The normalization constant corresponds to the probability of this particular projection which is
\begin{equation}\small
    p(\ket{\phi}) = \left|\bra{P}_{ab}\ket{\psi}_{AaBb}\right|^2 = \left|\frac{1}{2^5\sqrt{2}}\right|^2 = \frac{1}{2^{11}}.
\end{equation} 
This is the success probability of one particular detection pattern, but the protocol succeeds for all measurement outcomes where the photons are in different time bins. Depending on the specific detection pattern, different maximally entangled states will be prepared between Alice and Bob. However, the states are all equivalent up to local unitaries. A single target state like the one in Eq.~\eqref{eq:9} can thus always be achieved by performing a local correction on either Alice or Bob's qudit, dictated by the detection pattern. For the input state in Eq.~\eqref{eq:5} each of these projections is equally likely, resulting in a total success probability of $\frac{1}{8}$.  
\par
In Supplemental Material~A and B, we derive the projectors of the ESA for general even dimension, $d$, and show that entanglement swapping succeeds with probability $p_{\text{suc}} = \frac{2}{d^2}$. The intuition behind this scaling is that for a successful outcome, all photons must occupy different time-bins. The terms in the auxiliary state already contain photons in distinct time-bins. Furthermore, each term in the auxiliary state is missing photons in two time-bins, namely ${y_j, z_j}$. This implies that the photon on Alice's side must occupy time-bin $y_j$, while the photon on Bob's side must occupy time-bin $z_j$, or vice versa. The probability that this happens is $\frac{2}{d^2}$, which accounts for the scaling of our protocol.
Note that when applying the projection to the auxiliary state in Eq.~\eqref{eq:8}, each term in the auxiliary state corresponds to two non-orthogonal terms in $\bra{P'}$. Thus, the projection of input photons $a$ and $b$ will be a state with a Schmidt rank of two times the number of terms in the auxiliary superposition and as a consequence our protocol only works in even dimensions to project into a state with a Schmidt rank that is a multiple of two. 
\begin{figure*}[p]
    \centering  
    \includegraphics[width=0.9\textwidth]{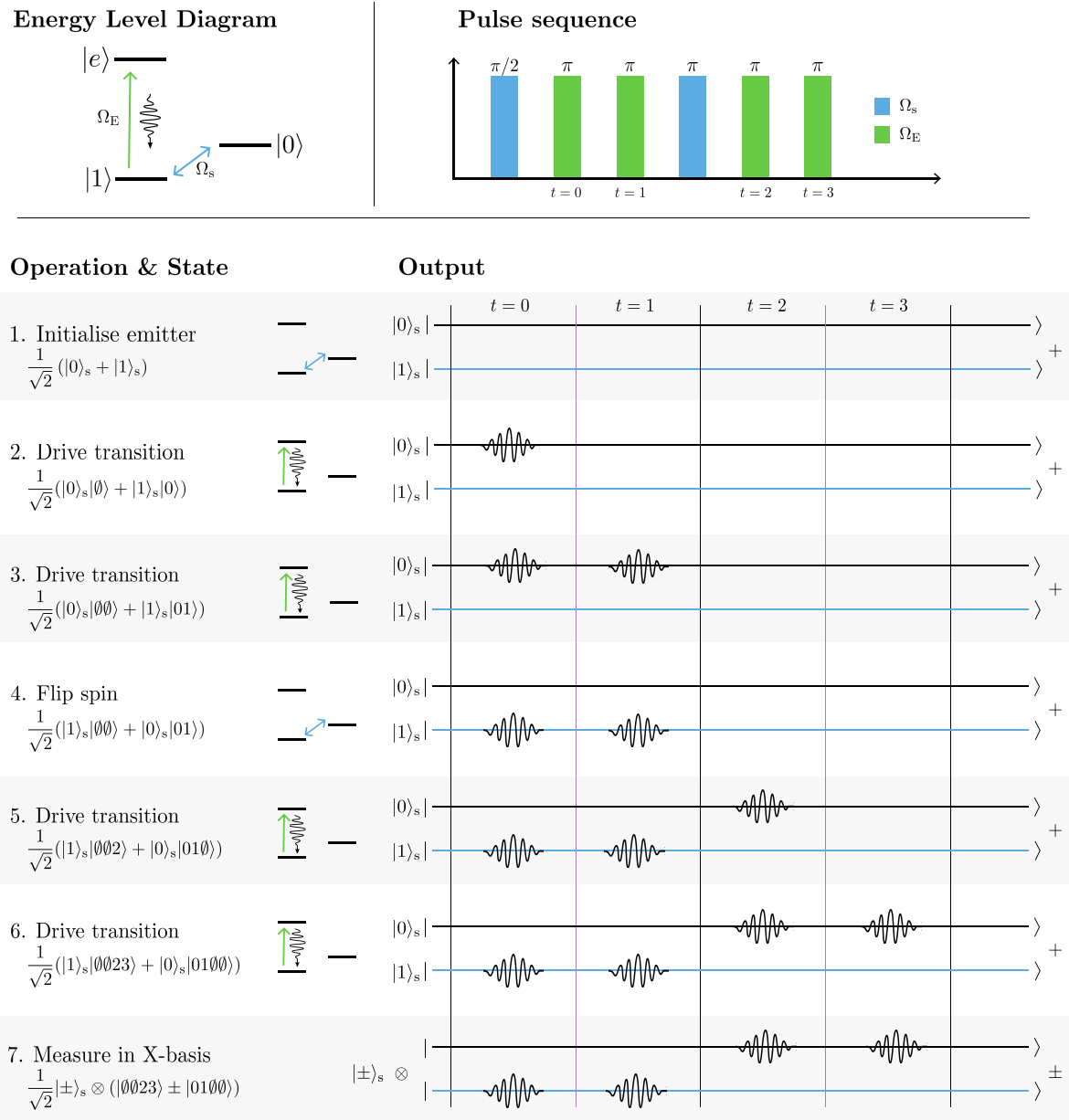}
    \caption{Top left: the energy level diagram of the quantum emitter that generates the state of auxiliary photons in the four-dimensional ESA. One of the ground states couples to an excited state that decays radiatively. Top right: the corresponding pulse sequence to generate the two auxiliary photon states. Bottom: diagram for generating the time-bin encoded auxiliary photons required for the four-dimensional ESA.}
    \label{fig:figure3}
\end{figure*}
\section{Preparation of the Auxiliary State}
The use of auxiliary photonic states in the form in Eq.~\eqref{eq:2} is key to achieving an efficient ESA, where the success probability only decreases quadratically with the dimension. 
In the four-dimensional ESA example, we can prepare the auxiliary state with one quantum emitter as shown in Fig.~\ref{fig:figure3}. 
\par
The emitter is initially prepared in an equal superposition $(\ket{0}_\text{s}+\ket{1}_\text{s})/\sqrt{2}$ of two ground states $\ket{0}_\text{s},\ket{1}_\text{s}$. We assume that there is a closed optical transition between the state $\ket{1}_\text{s}$ and an excited state $\ket{e}_\text{s}$. A short optical $\pi$-pulse can excite the population in the $\ket{1}_\text{s}$ state to the excited state followed by subsequent decay back to the $\ket{1}_\text{s}$ state with the emission of a photon. The emitter can be coupled to an optical cavity or waveguide for efficient collection of the photon but this is not a requirement. 
\begin{figure*}[tb]
    \centering
    \begin{subfigure}{0.45\textwidth}
        \includegraphics[width=\textwidth]{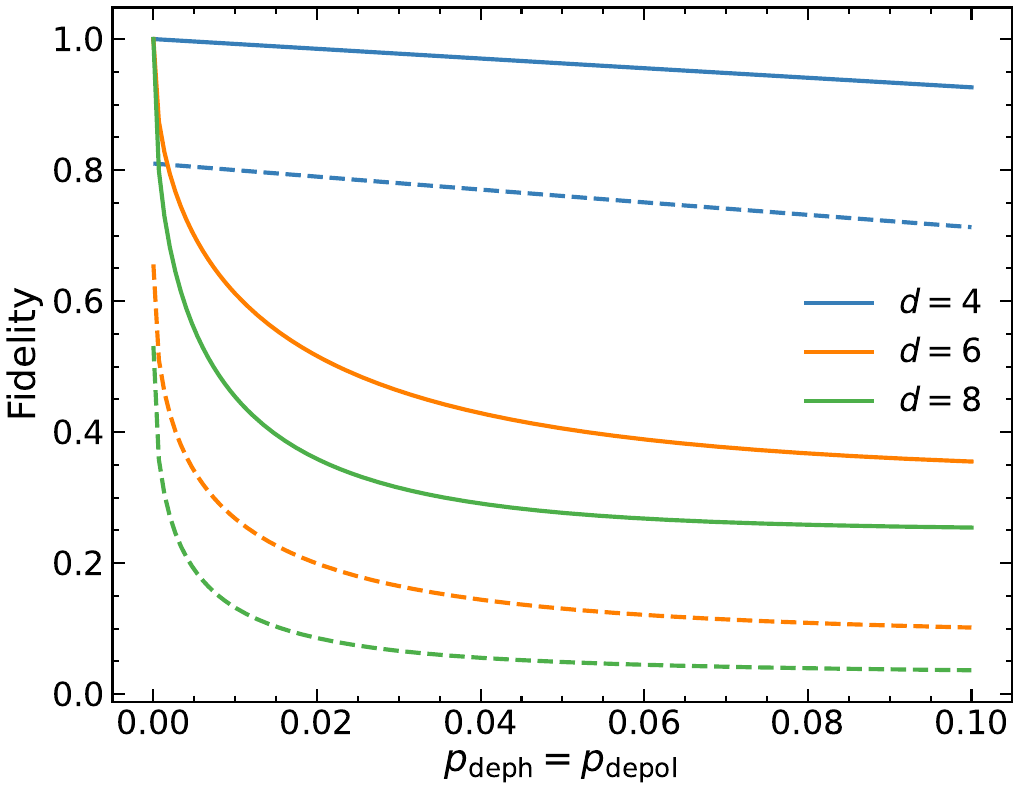}
    \end{subfigure}    
    \hspace{1.1cm}
    \begin{subfigure}{0.45\textwidth}
        \includegraphics[width=\textwidth]{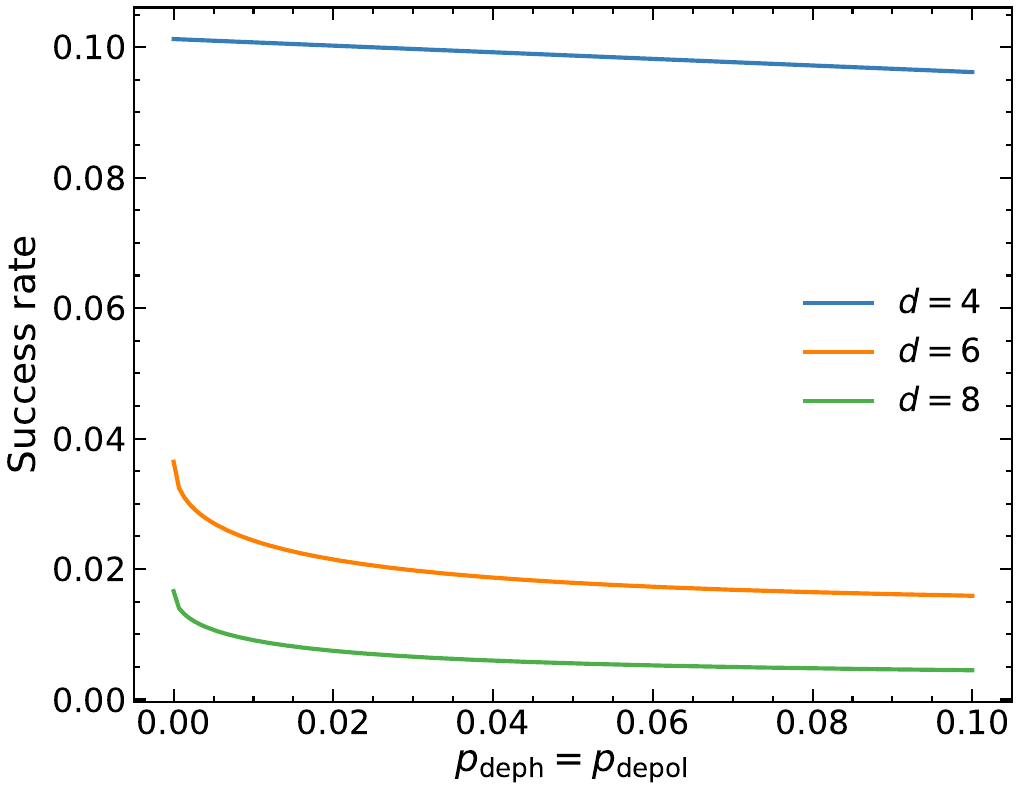}
    \end{subfigure}    
    \caption{Left: fidelity between the ideal auxiliary state and the imperfect auxiliary state (dashed lines) and the fidelity of the $(d-2)$-photon component of the imperfect auxiliary state (solid lines) in $4, 6$ and $8$ dimensions against dephasing $p_{\text{deph}}$ and depolarization $p_{\text{depol}}$. The probability of losing a photon in each emission step is $p_{\text{loss}} = 0.1$ for all dimensions. Right: success rate of the protocol with imperfect auxiliary state.}
    \label{fig:figure4}
\end{figure*}
\par
The proposal to prepare the required auxiliary multi-qudit state, illustrated by the pulse sequence and overviewed at the top of Fig.~\ref{fig:figure3}, is the following. We start by exciting the emitter, which is prepared in an equal superposition of the ground states. The emitter subsequently decays and emits one photon corresponding to the bright state in the ideal setting. We will first go through the scheme without considering experimental imperfections and consider a more realistic setting in the next section. We start by emitting two photons in subsequent time bins:
\begin{equation}
    \frac{1}{\sqrt{2}}\left(\ket{0}_\text{s}+ \ct_0\ct_1\ket{1}_\text{s}\right)\kvac.
\end{equation}
Here, the first ket indicates the spin qubit and the second ket the state of the photonic qudit, where $\ct_i$ is the creation operator of a photon in time bin $i$. Next, the spin is flipped

\begin{equation}
    \frac{1}{\sqrt{2}}\left(\ct_0\ct_1\ket{0}_\text{s} + \ket{1}_\text{s}\right)\kvac ,
\end{equation}
and we emit the next two photons:
\begin{equation}
    \frac{1}{\sqrt{2}}\left(\ct_0\ct_1\ket{0}_\text{s} + \ct_2\ct_3\ket{1}_\text{s}\right)\kvac.
\end{equation}
The photons are decoupled from the spin qubit by measuring the emitter in the $X$-basis. Depending on the measurement outcome ($\ket{\pm}=(\ket{0}\pm\ket{1})/\sqrt{2}$), the relative phase in the prepared state varies $\frac{1}{\sqrt{2}}(\ct_0\ct_1 \pm \ct_2\ct_3)\kvac$. This state has the desired shape from Eq.~\eqref{eq:4}. When the ESA is applied with this auxiliary state, the relative phase only leads to phase differences between the final state projections. Different auxiliary states can be prepared by applying a different sequence of X-gates and photon emissions.  
\par
The above procedure can be generalized to prepare the auxiliary states for arbitrary even dimensions. With a control register that contains $\lceil\log_2\left(\frac{d}{2}\right)\rceil$ qubits, using multi-qubit controlled gates, one can create the required state with $d-2$ photonic qudits and Schmidt rank $\frac{d}{2}$. The control qubits are prepared in an equal superposition of $d/2$ different states. By applying a series of multi-qubit controlled operations, the state of the spin qubit is flipped controlled on a specific term in the superposition of control qubits allowing for the emission of a photon conditioned on the state of the qubit register. Thus, each of the $d/2$ superposition states of the qubit register corresponds to one mode in the superposition of the auxiliary multi-qudit state. We refer the reader to Supplemental Material~C for further details.

\section{Error Analysis}
For near-term experiments, the operations of our protocol will be subject to imperfections, mainly during the generation of the auxiliary state and in the larger optical circuit. We analyzed the quality of the auxiliary state in $4$, $6$ and $8$ dimensions for the following experimental imperfections: (1) imperfect gate operations in the preparation of the auxiliary photons, (2) dephasing due to phase instability in the optical circuit including e.g. the QFT interferometer and (3) loss of photons and imperfect photon emission. 
\par
To keep the discussion general, we do not consider a specific platform for the control qubits or emitter, but rather model faulty gates as a depolarizing channel with depolarization probability $p_{\text{depol}}$ per unitary. Consequently, imperfect gate operations will cause photons to be emitted in incorrect time-bins or not at all as well as general dephasing of the auxiliary state.  
\par
While the photons pass through the optical fibers and the QFT circuit, optical instability can cause dephasing of the auxiliary state. We model this as a dephasing channel that acts on all auxiliary photons collectively. 
Furthermore, since the number of photons scales linearly with the dimension of the protocol, the probability of losing photons increases exponentially with dimension. Loss of at least one photon causes the protocol to fail since the detectors herald a photon loss as fewer than $d$ photons will be measured in the absence of detector dark counts. Assuming negligible dark count probabilities, photon loss will thus only decrease the success rate and not the fidelity of the scheme. Further details of the error model are provided in Supplemental Material~D.
\par
The fidelity of the auxiliary state is shown in Fig.~\ref{fig:figure4}. We vary the dephasing error and depolarizing error together and assume the loss probability per photon to be $p_{\text{loss}} = 0.1$. Since we consider the auxiliary state's fidelity, the total protocol's fidelity will be higher as the part of the auxiliary density matrix with less or more than $d$ photons can be heralded away at the detectors, trading higher fidelities for a decrease in success rates. Thus, we also calculate the fidelity of the part of the auxiliary state with exactly $d-2$ photons, which significantly improves the performance. We also show the total efficiency of the protocol in Fig.~\ref{fig:figure4}: the performance in $4$ dimensions is substantially better than the higher dimensions, which is to be expected since the auxiliary state is generated with a single emitter and, exceptionally, does not require a control register, as shown in Fig.~\ref{fig:figure3}. In higher dimensions, the generation of the auxiliary state becomes more involved, requiring more photons to be emitted and more multi-qubit operations controlled on a qubit register. We indeed see that the effect of errors is more prominent in higher dimensions. Nonetheless, the generation in $4$ dimensions is very robust against errors, which is promising for near-term implementations of the ESA.

\section{Conclusion and Discussions}
In summary, we have outlined an efficient linear-optics implementation of a high-dimensional entangled state analyzer in even dimensions. Our scheme uses the minimal number of required auxiliary photons and has a success probability that decreases only quadratically with the dimension. Importantly, we achieve this with auxiliary states with a Schmidt rank that increases only linearly with the dimension and we outline how the states can be generated deterministically with a single quantum emitter coupled to a small (logarithmic in dimension) qubit register.  
\par
Our work significantly relaxes the experimental requirement for a high-dimensional ESA and points to the use of quantum emitters for the efficient preparation of auxiliary states for linear-optics, high-dimensional photonics quantum processing. As outlined in this paper, the auxiliary state of a four-dimensional ESA succeeding with probability 1/8 can be generated from a single quantum emitter, paving the way for near-term experimental demonstrations using atomic~\cite{Thomas2022} or solid-state emitters~\cite{Knall2022,Meng2023}. 
\par
We investigated the performance in a faulty setting and showed that the success probability and fidelity are robust against errors, especially for $d=4$. We note that the auxiliary state of the four-dimensional ESA could also be prepared from an SPDC source in a probabilistic manner using a delay line for one of the emitted photons. While such an approach does not scale to higher dimensions, it may be relevant for near-term experiments with no access to a quantum emitter. Another approach is to use high-dimensional quantum emitters for the generation of the auxiliary state~\cite{Raissi2024}. This would eliminate the need for a qubit register and is suited for e.g. atomic emitters~\cite{Ringbauer2022,Thomas2022}. While the success probability of our scheme is on par with other, more complicated schemes in the literature, we are not ruling out that even better performance could potentially be obtained in the future. Investigation of this as well as the extension of our scheme to odd dimensions is left for future work.  

\begin{acknowledgments}
J.B. acknowledges support from The AWS Quantum Discovery Fund at the Harvard Quantum Initiative. L.M. was  supported by the Netherlands Organisation for Scientific Research (NWO/OCW), as part of the Quantum Software Consortium program (project number 024.003.037 / 3368).
 This work is supported by the Dutch National Growth Fund (NGF), as part of the Quantum Delta NL programme.

\end{acknowledgments}
\bibliographystyle{unsrtnat}
\bibliography{apssamp.bib}

\section*{Supplemental Material}
\subsection*{A. ESA in Arbitrary Even Dimensions} \label{app:A}
Here, we derive the general projections performed by our ESA scheme in arbitrary even dimension $d$ as shown in Fig.~\ref{fig:figure1}.
We use the same notation of $a_{j,k}$ and $y_j$ and $z_j$ as in the main text (see Eq.~\eqref{eq:2} and below). The general form of the four-dimensional projection from Eq.~\eqref{eq:6} in arbitrary even dimension and arbitrary detection pattern is 

\begin{widetext}
    \begin{equation}\label{eq:14}\footnotesize
        \begin{split}
            &\bra{P'} = \bvac\an_{0, D_0}\an_{1, D_1}\an_{2, D_2}\ldots\an_{d-1, D_{d-1}}\;U_{\textrm{QFT}} = \frac{1}{d^{d/2}} \bvac\left(\an_{0, a} + \omega^{D_0}\an_{0, b} + \omega^{2D_0}\an_{0, x_0} + \omega^{3D_0}\an_{0, x_1} + \ldots + \omega^{(d-1)D_0}\an_{0, x_{d-3}}\right)\\
            &\otimes\left(\an_{1, a} + \omega^{D_1}\an_{1, b} + \omega^{2D_1}\an_{1, x_0} + \omega^{3D_1}\an_{1, x_1} + \ldots + \omega^{(d-1)D_1}\an_{1, x_{d-3}}\right)\otimes\;\ldots\;\\
            &\otimes\left(\an_{d-1, a} + \omega^{D_{d-1}}\an_{d-1, b} + \omega^{2D_{d-1}}\an_{d-1, x_0} + \omega^{3D_{d-1}}\an_{d-1, x_1} + \ldots + \omega^{(d-1)D_{d-1}}\an_{d-1, x_{d-3}}\right)\\
            &= \frac{1}{d^{d/2}} \bvac\bigg[\omega^{D_1}\an_{0, a}\an_{1, b} + \omega^{D_0}\an_{1, a}\an_{0, b}\bigg]\an_{2, x_0}\an_{3, x_0}\an_{4, x_0}\dots\an_{d-1, x_0}\;\cdot\;\omega^{2D_2 + 2D_3 + \ldots + 2D_{d-1}}\\
            &\hspace{42.5pt}+ \bigg[\omega^{D_2}\an_{0, a}\an_{2, b} + \omega^{D_0}\an_{2, a}\an_{0, b}\bigg]\an_{1, x_0}\an_{3, x_0}\an_{4, x_0}\dots\an_{d-1, x_0}\;\cdot\;\omega^{2D_1 + 2D_3 + \ldots + 2D_{d-1}}\\
            &\hspace{20pt}+ \ldots +  \bigg[\omega^{D_{d-1}}\an_{d-2, a}\an_{d-1, b} + \omega^{D_{d-2}}\an_{d-1, a}\an_{d-2, b}\bigg]\an_{0, x_0}\an_{1, x_0}\an_{2, x_0}\dots\an_{d-3}\;\cdot\;\omega^{2D_0 + 2D_1 + \ldots + 2D_{d-3}}\\
            &= \frac{1}{d^{d/2}}\sum_{(t_0, t_1, \ldots, t_{d-1})}\bvac\an_{t_0, a}\an_{t_1, b}\an_{t_2, x_0}\an_{t_3, x_0}\ldots\an_{t_{d-1}, x_0}\;\cdot\;\omega^{D_{t_1} + 2D_{t_2} + 2D_{t_3} + \ldots + 2D_{t_{d-1}}}\\
            &\quad\quad\text{for}\quad(t_0, t_1, \ldots t_{d-1})\in P[0, 1, ..., d-1].
        \end{split}
    \end{equation}
\end{widetext}
\normalsize
Here $\omega$ is defined as $\omega = e^{\frac{2\pi i}{d}}$. $D_{i}\in \{0, 1, \ldots d-1\}$ represents the output port at which a photon in time-bin $i$ is measured. The expression above is a general expression that holds for every possible detection pattern where the photons are measured in different time bins. The projection describes the sum of all permutations of photons in different time bins, where each permutation term has a unique phase that depends on the specific detection pattern of the photons. From this expression, it is clear that every mode in the auxiliary state can lead to a successful detection pattern corresponding to two specific terms in the joint state of photons $a$ and $b$. Similar to Eq.~\eqref{eq:6}, we have omitted terms with more than one photon in spatial mode $a$ and $b$ and terms with photons in spatial modes $x_1, x_2\ldots x_{d-3}$ since these terms have zero overlap with the input state.
\par
Next, we use the last expression from Eq.~\eqref{eq:14} to project on the auxiliary input state to obtain the projection of the input modes $a$ and $b$ from the ESA:

\begin{equation}\label{eq:15}\footnotesize
    \begin{split}
        \braket{P'|\text{aux}} &= \frac{1}{\sqrt{d^d\cdot\frac{d}{2}}}\sum_{j=0}^{\frac{d}{2}-1}\omega^f\bvac\big(\omega^{D_{y_j}}\an_{z_j, a}\an_{y_j, b} \\
        &\hspace{3.2cm}+ \omega^{D_{z_j}}\an_{y_j, a}\an_{z_j, b}\big)\\
        &\text{where}\;f=2D_{\boldsymbol{\tau_j}[0]} + 2D_{\boldsymbol{\tau_j}[1]} + \ldots + 2D_{\boldsymbol{\tau_j}[d-3]}.
    \end{split}
\end{equation}\normalsize
\noindent Here the $\tau_{j}[k]$ correspond to the terms in the state of the auxiliary photons from Eq.~\eqref{eq:2}. Choosing specific $\tau_{j}[k]$ under the specified constraints defines the auxiliary state. As the constraints ensure that two time-bins ($y_j,z_j$) are excluded in each term of the auxiliary state, the underlying idea of the protocol is that we have tailored the auxiliary state such that in the projection of the photons $a$ and $b$, we project into an equal superposition of $d$ orthogonal modes, which is a maximally entangled state. As the phases in the projection depend on the detection pattern, we project onto different entangled states for different detection patterns. There are $d^d$ successful detection patterns meaning that different detection patterns do not result in projections onto orthogonal sets of maximally entangled states as is also clear from Eq.~\eqref{eq:14}. For entanglement swapping, the same output state can, however, always be output by means of local unitaries dictated by the detection pattern, as we show below.

\subsection*{B. High-Dimensional Entanglement Swapping}
Here we generalize the 4-dimensional entanglement swapping procedure from the main text to arbitrary even dimensions. The initial state of Alice and Bob is now:
\begin{equation}\label{eq:18}
     \ket{\psi}_{ABab} = \frac{1}{d}\sum_{i = 0}^{d - 1}\ct_{i, a}\kvac\ket{i}_A\otimes\sum_{j = 0}^{d - 1}\ct_{j, b}\kvac\ket{j}_B\;\;.
\end{equation}
If the ESA is successful, we can apply Eq.~\eqref{eq:15} to the initial state Eq.~\eqref{eq:18} to see that the state is projected into:

\begin{equation}\footnotesize
    \begin{split}
        \ket{\Psi}_{AB} &= \frac{1}{d\sqrt{d^d\cdot\frac{d}{2}}}\sum_{j = 0}^{\frac{d}{2} - 1}\omega^f\big(\omega^{D_{z_j}}\ket{y_j}_A\ket{z_j}_{B} \\
        &\hspace{2.6cm}+\omega^{D_{y_j}}\ket{z_j}_A\ket{y_j}_{B}\big).
    \end{split}
\end{equation}
Next, we apply the following unitary operation to the qudit register of Bob:

\begin{equation}\label{eq:16}\small
   \begin{split}
       &U(D_0, D_1, ..., D_{d-1}) =\\
       &\sum_{j = 0}^{\frac{d}{2} - 1}\omega^{-f}\big(\omega^{-D_{y_j}}\ket{y_j}\bra{y_j} + \omega^{-D_{z_j}}\ket{z_j}\bra{z_j}\big).
   \end{split}
\end{equation}
\normalsize
This results in the following state shared by Alice and Bob.

\begin{equation}\footnotesize
    \begin{split}
        \ket{\Psi}_{AB} = \frac{1}{d\sqrt{d^d\cdot\frac{d}{2}}}\sum_{j = 0}^{\frac{d}{2} - 1}\left(\ket{y_j}_A\ket{z_j}_{B} + \ket{z_j}_A\ket{y_j}_{B}\right).
    \end{split}
\end{equation}
\normalsize
Next we look at the output state for a specific choice of the auxiliary state, for example choosing the pairs $\{y_j,\;z_j\}$ consecutively: $(0, 1), (2, 3), \ldots, (d-2, d-1)$.
\begin{equation}\footnotesize
        \ket{\Psi}_{AB} = \frac{1}{d\sqrt{d^d\cdot\frac{d}{2}}}\sum_{j = 0,\; j\;\textrm{even}}^{\frac{d}{2} - 1}\left(\ket{j}_A\ket{(j+1)}_B + \ket{j+1}_A\ket{j}_B\right).
\end{equation}
\normalsize
The probability of measuring one successful detection pattern is:
\begin{equation}
    \left|\bra{\Psi}_{m}\ket{\psi}_{AB}\right|^2 = \left|\frac{1}{\sqrt{d}}\cdot\frac{1}{d\sqrt{d^d}\sqrt{\frac{d}{2}}}\cdot d\right|^2 = \frac{2}{d^2\cdot d^d}.
\end{equation} 
As long as all photons are measured in different time-bins, the combination of output ports is irrelevant as we showed in the previous section. Thus, each photon can be detected in $d$ output ports, and the protocol will succeed. There are $d$ photons and $d$ photon detectors, thus, $d^d$ detection patterns lead to success. The total success probability is a polynomial function of dimension:

\begin{equation}\label{eq:22}
    p_{\textrm{ES}} = \left|\bra{P}_{ab}\ket{\psi}_{ABab}\right|^2\cdot d^d = \frac{2}{d^2}\;\;.
\end{equation}

\subsection*{C. Auxiliary Photon State}
We described the shape of the auxiliary state in Eq.~\eqref{eq:2} which is equivalent to the following formulation.

\begin{equation}\label{eq:23}\small
    \begin{split}
        \ket{\textrm{aux}}& = \frac{1}{\sqrt{d/2}}\sum_{j = 0}^{\frac{d}{2} - 1}\bigotimes_{k = 0}^{d-3}\ct_{\boldsymbol{\tau_j}[k]}\kvac\\
        = &\frac{1}{\sqrt{\frac{d}{2}}}\big(\ct_{\boldsymbol{\tau_0}[0]}\ct_{\boldsymbol{\tau_0}[1]}\hdots\; \ct_{\boldsymbol{\tau_0}[d-3]}\kvac \\
        &+ \ct_{\boldsymbol{\tau_1}[0]}\ct_{\boldsymbol{\tau_1}[1]}\hdots \;\ct_{\boldsymbol{\tau_1}[d-3]}\kvac\\
        &+ \hdots\; \\
        &+ \ct_{\boldsymbol{\tau_{(d/2)) - 1}}[0]}\ct_{\boldsymbol{\tau_{(d/2) - 1}}[1]}\hdots\; \ct_{\boldsymbol{\tau_{(d/2) - 1}}[d-3]}\kvac\big).
    \end{split}
\end{equation}
with
\begin{equation}\label{eq:24}
    \begin{split}
            &(\boldsymbol{\tau_j}[0], \boldsymbol{\tau_j}[1], ..., \boldsymbol{\tau_j}[d-3]) \in P[0, 1, ..., d-1]\\
            &(\boldsymbol{\tau_j}[0], \boldsymbol{\tau_j}[0], ..., \boldsymbol{\tau_j}[d-3])\neq \{y_j, z_j\}\\
            &(y_j, z_j)\in\{0, 1, ..., d-1\}^2\\
            & y_i\neq y_j\;\land\; z_i\neq z_j\;\land\; y_i \neq z_j \quad\forall\; i, j\;\;.
    \end{split}
\end{equation}
Where the creation operator $\ct_{\boldsymbol{\tau_j}[k]}$ corresponds to a photon created in time-bin $\boldsymbol{\tau_j}[k]$ in spatial mode $x_0$ and the variables $\boldsymbol{\tau_j}[k], y_j, z_j$ should satisfy the constraints from Eq.~\eqref{eq:24}. To prepare the auxiliary state for the protocol in arbitrary even dimensions, we use a register with a quantum emitter that has $(d/2) + 1$ energy levels where each state selectively emits a photon. This can be realized with a single three state emitter system, as shown in Fig.~\ref{fig:figure3}, coupled to a $\lceil \log_2(d/2)\rceil$ qubit processor. We use this register as a control register to flip the emitter state by applying gates conditioned on a specific state of the qubit register. In contrast to the example in $4$ dimensions in the main text, we now initialize the quantum emitter in the bright state. This protocol can be used in specific dimensions where $d = 2^{x+1}$ for integer $x$ and in doing so, we reduce the number of required multi-controlled bit flips to the emitter. In arbitrary even dimension, the emitter should start in the dark state and our protocol can be used in the same way but requiring more multi-controlled bit flips. The initial state is:

\begin{equation}\label{eq:25}\small
    \begin{split}
        \frac{1}{\sqrt{\frac{d}{2}}}\left(\ket{c_0} + \ket{c_1} + \ldots + \ket{c_{(d/2) - 2}} + \ket{c_{(d/2)-1}}\right)\otimes\ket{0}_\text{s}.
    \end{split}
\end{equation}
With $\ket{c_j}$ the state of the control qubits and $\{\ket{c_0}, \ket{c_1}, \ldots, \ket{c_{(d/2) - 1}}\}$ spanning a basis, for example the computational basis. In the protocol, the following operations are applied to flip and excite the quantum emitter twice controlled on one mode of the control register. 
\begin{enumerate}
    \item Flip the quantum emitter controlled on one of the modes $\ket{c_j}$ in the superposition of control qubits such that it is in the dark state corresponding to that mode.
    \item Excite the quantum emitter to emit a photon for all states except $\ket{c_j}$.
    \item Excite the quantum emitter in the next time-bin to emit a photon for all states except $\ket{c_j}$.
    \item Execute step $1$ again to flip the quantum emitter back to the bright state.
\end{enumerate}
We call this pulse sequence to emit two consecutive photons '$U^j_{\text{em}}$', where the superscript $j$ corresponds to a specific control mode $\ket{c_j}$.
\par
Now we have all the elements to build up the state of the auxiliary photons. We emit photons in time-bins $0$ and $1$ corresponding to all modes except $\ket{c_{\frac{d}{2} - 1}}$, so we start with operation $U^{\frac{d}{2} - 1}_{\text{em}}$. The state after this operation is:

\begin{equation}
    \begin{split}
        \sqrt{\frac{1}{d / 2}}\bigg(&\big(\ket{c_0} + \ket{c_1} + \;\ldots\; +\ket{c_{(d/2) - 2}}\big)\ct_0\ct_1\kvac\\
        &+ \ket{c_{(d/2) - 1}}\kvac\bigg)\otimes\ket{1}_\text{s}.
    \end{split}
\end{equation}
\\
The second kets indicate the auxiliary photons in spatial mode $x_0$. We emit the next photons controlled on all modes except $c_{\frac{d}{2} - 2}$ with operation $U^{\frac{d}{2} - 2}_{\text{em}}$:

\begin{equation}
    \begin{split}
        &\sqrt{\frac{1}{d / 2}}\bigg(\big(\ket{c_0} + \ket{c_1} + \ket{c_2} + \ldots\;\big)\ct_0\ct_1\ct_2\ct_3\kvac\\
        &+\ct_0\ct_1\ket{c_{(d/2) - 2}}\kvac + \ct_2\ct_3\ket{c_{(d/2) - 1}}\kvac\bigg)\otimes\ket{1}_\text{s}.
    \end{split}
\end{equation}
We continue emitting photons controlled on all modes except for one with operations $U^{\frac{d}{2} - 3}_{\text{em}}\ldots U^{1}_{\text{em}}U^{0}_{\text{em}}$. After the last operation, the state is:

\begin{equation}\label{eq:29}\small
    \begin{split}
        \sqrt{\frac{1}{d / 2}}&\bigg(\ct_0\ct_1\ct_2\ct_3\hdots\ct_{d-3}\ket{c_0}\kvac\\
        &+ \ct_0\ct_1\ct_2\ct_3\hdots\ct_{d-5}\ct_{d-2}\ct_{d-1}\ket{c_1}\kvac\\
        &+ \ct_0\ct_1\hdots\ct_{d-7}\ct_{d-4}\ct_{d-3}\ct_{d-2}\ct_{d-1}\ket{c_2}\kvac \\
        &+\;\ldots\\
        &+ \ct_0\ct_1\ct_4\ct_5\hdots\ct_{d-1}\ket{c_{(d/2) - 2}}\kvac\\
        &+ \ct_2\ct_3\ct_4\ct_5\hdots\ct_{d-1}\ket{c_{(d/2) - 1}}\kvac\bigg)\otimes\ket{1}_\text{s}.
    \end{split}
\end{equation}
Where each photonic mode corresponding to a mode from the control qubit contains photons in all modes except two in a unique way: the photonic mode of $c_0$ does not contain photons in time-bin '$d-2$' and '$d-1$', $c_1$ does not contain photons in time-bin '$d-4$' and '$d-3$' and so on. To remove the correlation of the photonic qudit states with the control register, we finally measure the control register in the Fourier basis. The outcomes determine the phases in the auxiliary state, which is still a suitable state of the form from Eq.~\eqref{eq:2}. The measurement outcome can be used to determine which entangled state is generated exactly. 
\par In Eq.~\eqref{eq:29} we have prepared a specific instance of the auxiliary state. The same procedure can be applied to prepare different auxiliary states. In each time-bin, photons need to be emitted conditional on certain modes in the control register. We need to apply multicontrolled operations from these modes to the emitter, such that the emitter is in the bright state correlated to these modes. Next, the emitter is excited and a photon is emitted correlated with these modes. Applying the same multicontrolled operation returns the emitter to the dark state and the process is repeated for the next time-bin. Measuring the control register in the Fourier basis projects the photons into the desired auxiliary state.

\subsection*{D. Decoherence and Photon Loss}
In dimensions $6$ and $8$, we model the faulty multi-qubit controlled operations in the generation of auxiliary photons as single qubit depolarizing channels acting on each qubit following a perfect gate operation. In $4$ dimensions this is how we model the imperfect $X$-gate of the emitter. The Krauss operators are:

\begin{equation}
    \begin{split}
        K_0 &= \sqrt{1 - \tfrac{3p_{\text{dp}}}{4}}\;\begin{bmatrix}
            1 & 0\\
            0 & 1
        \end{bmatrix}\\
        K_1 &= \sqrt{\tfrac{p_{\text{dp}}}{4}}\;\begin{bmatrix}
            0 & 1\\
            1 & 0
        \end{bmatrix}\\
        K_2 &= \sqrt{\tfrac{p_{\text{dp}}}{4}}\;\begin{bmatrix}
            0 & -i\\
            i & 0
        \end{bmatrix}\\
        K_3 &= \sqrt{\tfrac{p_{\text{dp}}}{4}}\;\begin{bmatrix}
            1 & 0\\
            0 & -1
        \end{bmatrix}.\\
    \end{split}
\end{equation}
The depolarizing channel acting on density matrix $\rho$ that describes a single qubit is modeled as:

\begin{equation}
    \tilde{\rho} = \sum_{i = 0}^3K_i\rho K_i^\dag.
\end{equation}
In dimension $4$, we depolarize the emitter. In dimensions $6$ and $8$ we apply the single qubit depolarization to both qubits in the control register and the emitter. We model the dephasing of auxiliary photons due to the instability of the linear optics circuit and fibers as a collective dephasing. The dephasing channel acting on density matrix $\rho$ of size $x$ x $x$ is given by:
\begin{equation}
    \tilde{\rho} = (1-p_{\text{deph}})\rho + p_{\text{deph}}\text{diag}(\rho_{00}, \rho_{11}, \ldots \rho_{xx}).
\end{equation}
Each photon can be lost somewhere in the circuit and we model this as a general loss probability $p_{\text{loss}}$. For example, if we want to emit a photon in time-bin $i$ conditional on the mode of the control qubits $c_j$ and for all other modes $c_i$ where $i\neq j$ not emit a photon, the faulty operation looks like:
\begin{equation}
    \begin{split}
        &\ket{c_j} \rightarrow \sqrt{1-p_{\text{loss}}}\ct_i\ket{c_j}\kvac\\
        &\ket{c_i} \rightarrow \ket{c_i}\kvac.
    \end{split}
\end{equation}
This leads to an unnormalized density matrix which effectively is equivalent to assuming a fidelity of zero when no photons are emitted when they should be emitted. 
\par
As discussed previously, the terms with fewer photons can be heralded away at the detection. To see the effect of this heralding we calculate the fidelity of the auxiliary state conditional on the number of photons being equal to $d-2$, the number of auxiliary photons that should be there in the ideal case. In the presence of detector dark counts, auxiliary states with less than $d-2$ photons could lead to a false success due to a dark count. In addition, events where more than $d-2$ photons are emitted could also lead to a false success if the photon detectors are not number resolving. As these errors are arguably higher order compared to the photon loss and the depolarizing errors considered here, we do not include them in our model.

\end{document}